# Analysis of training and seed bias in small molecules generated with a conditional graph-based variational autoencoder – Insights for practical AI-driven molecule generation

Seung-gu Kang*, Joseph A. Morrone†, Jeffrey K. Weber† and Wendy D. Cornell

Computational Biology Center, IBM Thomas J. Watson Research
1101 Kitchawan Road, Yorktown Heights, NY 10594, USA

**Abstract**

The application of deep learning to generative molecule design has shown early promise for accelerating lead series development. However, questions remain concerning how factors like training, dataset, and seed bias impact the technology's utility to medicinal and computational chemists. In this work, we analyze the impact of seed and training bias on the output of an activity-conditioned graph-based variational autoencoder (VAE). Leveraging a massive, labeled dataset corresponding to the dopamine D2 receptor, our graph-based generative model is shown to excel in producing desired conditioned activities and favorable unconditioned physical properties in generated molecules. We implement an activity swapping method that allows for the activation, deactivation, or retention of activity of molecular seeds, and we apply independent deep learning classifiers to verify the generative results. Overall, we uncover relationships between noise, molecular seeds, and training set selection across a range of latent-space sampling procedures, providing important insights for practical AI-driven molecule generation.

**Introduction**

The design of small molecule drugs is a challenging problem owing to requirements for target activity and selectivity as well as limitations imposed by synthetic access and tractability.[1–3] Medicinal chemists apply advanced pattern recognition skills to guide structure-activity analysis and design, and these efforts can be augmented by statistical methods and physics-based models such as docking. Historically, classification or regression predictions on existing or proposed molecules have been the dominant approach[4,5] but *de novo* design methods have also been pursued.[6,7]

The recent advent of deep learning has opened the door to generative AI-based molecule design. A myriad of publications centered on design of small molecule drugs optimized for activity against specific protein targets have appeared in the literature over the past five years.[8–14] The majority of such efforts have employed simple, 1D string-based representations of ligands as a starting point for structural encoding, almost always using the SMILES[15] language. These string-based descriptions of molecules have been attractive as compact textual representations that afford direct application of deep learning methods developed in the natural language processing space.[16] However, recent work has identified pitfalls with using SMILES strings in the area of generative molecular applications, particularly with respect to the considerable degeneracy encountered in representation space.[17–19] Typically, many distinct SMILES strings can correctly describe the same molecule, complicating pattern matching in the context of evaluating reconstruction loss functions that are central to training variational autoencoders (VAE). Some researchers have proposed extensive SMILES enumeration protocols to resolve this issue of string degeneracy[17,18]; others have proposed abandoning SMILES entirely in favor of more syntactically complex representations (related to SMARTS, for example) that at least partially ameliorate the problem.[20,21]

An alternative representation approach harkens to an idea more grounded in chemistry: molecules are not one-dimensional entities. Ignoring, for now, the effects of ligand conformational diversity and stereochemistry, molecules can be projected in two dimensions as networks of atoms connected by their respective bonds. Two-dimensional graphs are thus natural and attractive representations for deep learning frameworks, and are further buoyed by methods like graph convolutions that allow for abstraction of local chemical structure at multiple resolutions. Deep learning classification based on 2D graph convolutions has become popular in Quantitative Structure Activity Relationship (QSAR) modeling,[22–24] and similar examples of graph-based models have appeared in the literature in the context of generative applications.[13] While graphs are not completely immune to concerns over isomorphism at the level of reconstruction loss, graph theory suggests simple means of rectifying such issues. Throughout the generative small molecule design literature, and particularly with respect to graph-based VAEs, performance evaluation has focused on molecule validity, uniqueness, and physical properties alongside simple target activity predictions based on labeled data.[25-28] In many cases, however, typical machine learning protocols that are followed for classification tasks (like holding out independent training and validation sets) have often not been applied in generative cases, and sources of bias (related to mutual ligand similarity or other considerations) within training sets have not been extensively explored. The use of protein-ligand complex information in the field of virtual screening provides a cautionary tale for new deep learning methods in drug discovery: early promising data have now been determined to be the spurious results of ligand-similarity bias in commonly used training sets.[29-31] It is thus imperative to ensure that facets of training and latent space sampling bias are well understood in the generative small molecule modeling space.

In this work, we present a conditional graph-based generative modeling architecture that allows for the flexible specification of generated ligand activity. We apply this model to a massive, labeled dataset corresponding to ligand activity against the D2 dopamine receptor, and we assess the effects of various latent space training, conditioning, and sampling protocols. We explore whether and how molecules can be generated that correspond to specific conditions, and we further evaluate 1) if active molecules can be generated from direct(random) or seed-based latent space sampling *via* modulation of a binary activity condition, and 2) if seed molecules can be either "activated" or "deactivated" against our chosen target using a binary condition swap. We also assess how division of training and evaluation sets impact the predicted properties of generated molecules, and we investigate the degree to which innate molecule properties are transferred from training sets to generated distributions of molecules. Overall, we show that molecular seed and training set biases have considerable effects on latent space sampling and corresponding generated molecules, suggesting that such factors need to be accounted for in applications of AI-driven small molecule design.

## Methods
### Graph-based conditional variational autoencoder
Small molecules are here described as mathematical graphs composed of nodes and edges representing atoms and chemical bonds, respectively. In the present implementation, each node was represented by a vector containing the atom's chemical element type ($N_A = 10$, $A = \{a | a \in (C, N, O, F, P, S, Cl, Br, I, \emptyset)\}$) and, when the atom belonged to a ring, additional features describing ring characteristics (i.e., aliphatic or aromatic identity). Graph connectivity, as inspired by a past relational graph convolution network (R-GCN)[32], was represented using a tensor of stacked adjacency matrices, each constructed according to a given bond type ($N_B = 5$, $B = \{b | b \in (single, double, triple, aromatic, \emptyset)\}$). Since one of our design protocols involves selectively generating molecules to be either active or inactive against given target (with molecules either randomly generated from the latent space or centered around a designated seed molecule), the encoder was conditioned with a binary target activity vector component concatenated onto each node vector.

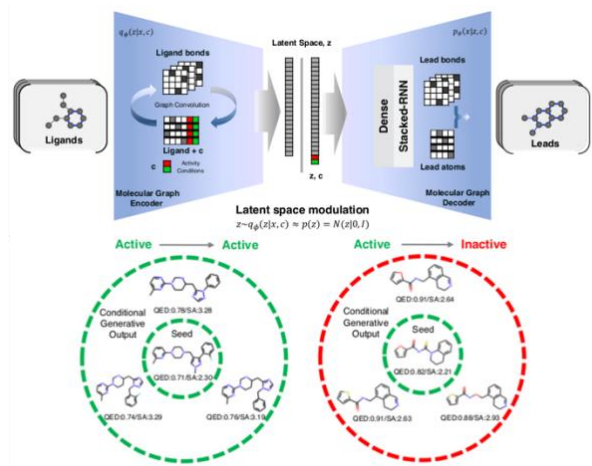

**Figure 1** | Architectural schematic of conditional-VAE based relational graph-convolutional neural network, presented above illustrations of our activity conditioning protocol.

**Molecular encoding with relational GCN** Molecular graphs were encoded using a graph convolutional network as a function of each bond type as follows[32]:

$$h_i^{(l+1)} = \sigma\left(\sum_{b \in B} \sum_{j \in n_i^b} \frac{1}{N_i^b} W_b^{(l)} h_j^{(l)} + W_s^{(l)} h_i^{(l)}\right)$$

where the $i$th hidden node of the $l$th hidden layer ($h_i^{(l)}$) is convolved with its neighbor nodes ($h_j^{(l)}$) after a dense transformation ($W_b^{(l)}$) over all available neighbor nodes ($n_i^b$), connected through a bond type, $b$, and normalized by the number of connected neighbors ($N_i^b$). Convolutional neurons are activated with the tanh function, $\sigma(\cdot)$, after adding a self-connection term ($W_s^{(l)} h_i^{(l)}$). In the current encoder, graph convolutions were carried out to a depth of two, with each convolutional operation preceded by two dense layers (128 and 64 dimensions). Following a 128-dimension dense transformation, nodes were aggregated by summing over an element-wise product of sigmoid and tanh activations, respectively, in order to produce an embedding invariant to atom order.

**Latent space sampling with activity conditioning** The graph convolutional embedding was projected onto variational parameters with mean ($\mu$) and standard deviation ($\sigma$) over a 32-dimensional vector that resulted from successive 128- and 64-dimensional dense neural layer operations. Latent vectors ($z'$) were sampled using Gaussian noise $\epsilon \sim N(\mathbf{0}, \mathbf{1})$: $z' = \mu + \epsilon\sigma$. As in the encoder, the binary target activity condition was concatenated with $z'$ to give the final latent embedding vector, $z$. By further conditioning at the embedding level, we enabled activity control over generated molecules at two levels: 1) starting from the encoder and based on molecular seeds, or 2) from randomly sampling and decoding latent space vectors (without seeds).

**RNN-assisted molecular graph decoder** Our molecular graph decoder was designed to generate molecular relational graphs, i.e., node vectors for atoms and multiple edge matrices for each bond type. In the present network, all nodes and edges were decoded in one shot. Latent embeddings were first passed through two fully connected dense layers (64 and 128 dimensions) with tanh activation functions and then expanded in dimension to meet a set maximal number of atoms, $N_{max}$, per generated molecule ($N_{max}$=40 atoms in our study). In order to enhance the quality of generated molecular graphs, a dynamic recurrent neural network (RNN) was added prior to the node and edge generative operations. A total of 512 LSTM cells were incorporated within the RNN. The RNN output was fed into the node, edge, and atomic feature prediction layers. Nodes and edges were generated using independent linear transformations with no activation to yield $N \times N_A$ and $N \times N \times N_B$ tensors, respectively, while each decoded atomic feature vector was obtained using two layers of dense transformations (8 and 4 dimensions) with tanh activation functions. Decoder performance (measured by the chemical validity of output molecular graphs) was optimized by modulating RNN hyperparameters and specific atomic feature types.

**Loss function and optimization** Providing an activity condition at input, our generative model was trained by maximizing the variational lower bound, L as defined by:

$$L = \mathbb{E}_{q_\phi(z|x,c)}(\log p_\theta(x|z,c)) - KL(q_\phi(z|x,c)\|p(z))$$

The first term in this equation represents the sum of reconstruction losses for node, edge and atomic features. Here, $p_\theta(x|z,c)$ is the probability distribution over molecular graphs generated from the decoder network ($\theta$) with activity condition, $c$, averaged over $q_\phi(z|x,c)$, the posterior distribution of the latent variables ($z$) approximated by the encoder network ($\phi$) with the same target activity condition ($c$). The second term in the loss function is the Kullback-Leibler (KL) divergence between the Gaussian prior, $p(z)$, and the posterior, $q_\phi(z|x,c)$. The KL divergence serves to regularize generated output. All our models were built based on the framework of the TensorFlow machine learning python library[33].

**Training Database and Model Training**
All models are built using high-throughput screening assay data from PubChem Assay ID 485358.[34] This dataset consists of 1780 active-labeled and 324800 inactive-labeled molecules with respect to their binding to the dopamine D2 receptor. Two separate datasets were curated for the purpose of training generative models: (1) a subset of the full D2 dataset containing all active molecules and half the inactive ones, and (2) a subset of the full D2 dataset containing half the active molecules and half the inactive molecules. The latter split facilitates a more independent assessment of our model using the held-back data (see below). Reserving independent validation data is not particularly common practice in generative chemistry fields, but choosing to do so allows for model evaluation less sensitive to molecular features that can be learned directly from the training data. Presently, each D2 dataset is randomly split into training and test components. Other data splitting strategies, including grouping ligands in the training set based on their similarity,[35] may be explored in future work. However, the present splitting scheme is sufficient for an initial demonstration of our methods.

All generative models were trained under the framework of the TensorFlow python library.[36] Model weight parameters were optimized with the Adam optimizer[37] with default learning rate = 0.001, $\beta_1$ = 0.9, and $\beta_2$ = 0.999 settings; weights were updated after every batch (each containing 256 data points) over 1,000 epochs. The KL contribution to the loss function was gradually increased as training proceeded, as often seen with $\beta$-VAEs.[38] Increasing $\beta$ over time allows the network to be more efficiently optimized in parameter space with respect to molecular graph validity during early epochs of optimization while growing more regularized at later stages of training.

**Validating activity of generated molecules**
In order to assess generated molecules, we built a ligand-based activity classification model for small-molecule binding to the D2 receptor. This deep learning QSAR (DL-QSAR) model uses atom types and the chemical bond network needed for input into a graph-convolutional architecture based on the "graphconv" module in DeepChem[23]. The output layer of this network is a binary softmax classifier. This architecture has been adapted for use in our in-house code[31] built with the TensorFlow library[36]. The hyperparameters of the model are the same as those employed in Ref. [31]

The DL-QSAR model is trained in two ways: (1) with all data from PubChem Assay ID 485358 and (2) with the half of the dataset held back from training the generative model. The model as trained in scenario (1) was used in initial assessment of our generative models, whereas the QSAR model from (2) produced the results presented in **Fig 4**. In the latter case, the DL-QSAR model trained on one-half the data was paired with the generative model trained on the other half of the data. Four independent training runs were undertaken for each scenario and the resultant predictions were averaged over all runs.

**Calculating Molecule Properties**
Molecular properties (molecular weight, QED and SA) of the training and generated data were computed with the RDKit cheminfomatics python library,[39] where QED and SA scores were obtained based on Bickerton et al.[40] and Ertl et al.[41]'s methods, respectively. Molecular similarities were calculated using the Dice similarity metric for Morgan circular fingerprints[42] of radius 2. Principal component analysis was performed with the PCA module provided by sklearn python machine-learning package.[43]

**Protocols for Molecule Generation**
Using our trained conditional VAEs, we generated molecules using two sets of protocols: 1) direct (random) sampling from the latent space, and 2) seed-based sampling based on designated input molecules. In the former case, the latent embedding was sampled directly from the Gaussian random noise accompanied with the desired activity condition. In the latter case, a seed molecule and corresponding activity condition were fed through the encoder to produce an embedding, which was then augmented with Gaussian noise. Depending on the specified activity condition, our generation protocols produced four output subgroups dubbed **rand0**, **rand1**, **seed0,** and **seed1**, where **0** and **1** represent inactive and active labels, respectively. Specifically, **seed0**(**seed1**) indicates a set of molecules generated by encoding inactive(active) molecules followed by decoding with an inactive(active) label, while **seed01**(**seed10**) by encoding inactive(active) seeds followed by decoding with active(inactive) activity label. We call this procedure "seed-swapping" throughout the rest of the text. For each of these conditional subgroups, we prepared four different sets of molecules at various distances from the variational mean with noise standard deviations 0.5, 1.0, 2.0, and 4.0 (indicated with s0.5, s1.0, s2.0, and s4.0, respectively).

## Results and Discussion
**Generative Model Can Learn Unconditioned Molecular Features**
Although modulation of biological activity is of particular interest in the context of molecular generation, physical properties are equally important, so we first investigated their outcomes when not explicitly conditioned in the network. To what extent were these other properties of the training data reflected in generated molecules? We surveyed three different aspects of generated molecular structure with respect to this question: i) molecular weight (MW) as an overall measure for the molecular size, ii) quantitative estimate of drug-likeness (QED) as a broad assessment for drug-like physical properties, and iii) synthetic accessibility (SA) as a gauge for synthetic feasibility.

**Molecular weight** In **Fig 2**, we show the MW distributions for the four groups of molecules defined by sampling method (random or seed-based) and activity condition (i.e., active or inactive). For the random sampling use cases, **rand0** and **rand1** (**Fig 2 A&B**), the generated molecules were found to have a smaller molecular weight than the training set, regardless of the conditioned activity. While the training set has a MW distribution peaked at 366(±75), for example, random sampling at s0.5 around the Gaussian prior mean (i.e., **μ=0**) yielded a Gaussian distribution with a peak around 327(±35). As molecules were sampled with progressively larger standard deviations, the MW distribution gradually broadened and shifted toward smaller values. Molecular size is thus inversely proportional to sampling distance from the prior mean in the random sampling scheme, suggesting that valid molecules tend to become smaller as one deviates from **μ=0**. This observation perhaps makes intuitive sense, as large, complex molecules are naturally difficult to generate at large distances from training data points. At low standard deviations (near **μ=0**), molecular sizes tend to more closely mirror the training distribution.

Seed-based generation (**seed0** and **seed1**) yielded dramatically different results from random sampling in terms of molecular weight. As shown in **Fig 2 C&D**, generated MW distributions are closely pinned to those of corresponding seed molecules. The distributions do vary somewhat as a function of standard deviation: at points, sampling at small standard deviations (e.g., s0.5) generates molecules closer in size to the underlying seed molecules than does sampling at larger standard deviations. However, the MW of generated molecules is strikingly coupled to the MW distribution of the seeds, regardless of the standard deviation. Virtually no band or population shifts are observed on the lower MW side of the seed-based distributions, in stark contrast to the random sampling case. These results highlight both potential benefits and limitations of the two sampling methods: molecules of relatively large, drug-like size are efficiently generated through the seed-based approach, but molecules that differ more from the training data are likely more accessible through random latent space sampling. The relatively small size of the latter leaves room for adding or tweaking groups to manually refine the new lead.

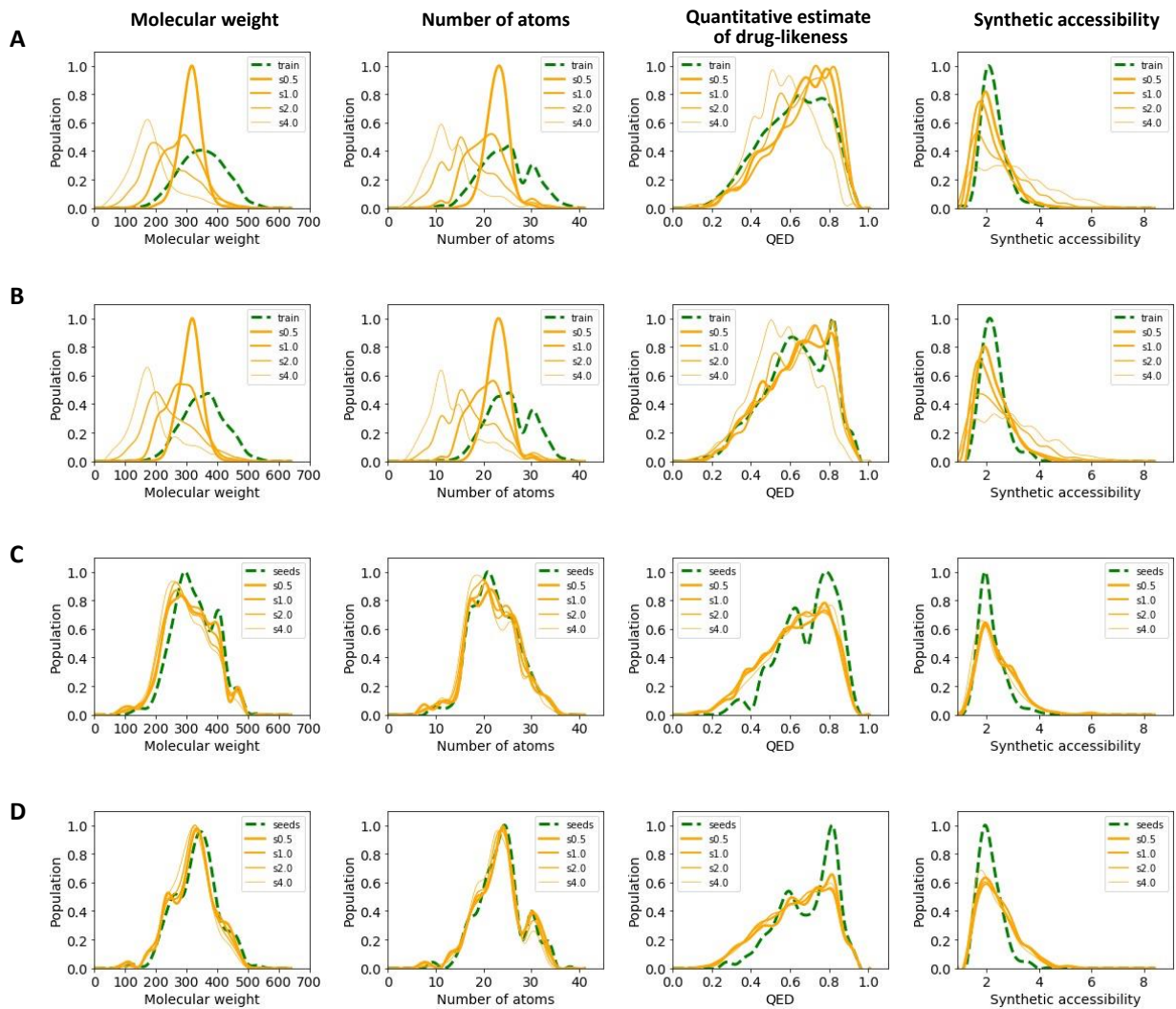

**Figure 2** | Molecular properties of generated molecules. **A** & **B**, random sampling with inactive and active conditions, respectively. **C** & **D**, seed-based sampling with inactive and active conditions on cognate seeds, respectively. Each column shows molecular weight, number of atoms, quantitative estimate of drug-likeness (QED) and synthetic accessibility from the left to the right.

**Quantitative Estimate of Drug-likeness** QED distributions for generated molecules illustrate the extent to which generic drug-like properties were captured in our generative modeling. Both active and inactive molecules in the training feature various subpeaks, with defined, separate peaks at intermediate and high QED values.

We compared molecules generated with random sampling in **Fig 2A&B**. Independent of activity condition, generated molecules exhibited similar QED distributions to the training sets at low standard deviation (s0.5 and s1.0), but the mean QED decayed toward ~0.5 at higher noise levels. Differences in distribution features (e.g., peak position or width) are not surprising considering the noisy nature of sampling from the Gaussian prior. However, particularly at lower noise levels, generated molecules generally resemble training molecules in terms of drug-likeness.

Molecules generated with the seed-based protocol exhibit a similar pattern in terms of QED: even at the highest noise level, the QED distribution over training molecules is largely maintained in generated examples, albeit with some smoothing and a small degree of bleeding into the low QED regime. While the desirable high QED peak could not be fully reproduced, the probability of sampling molecules between the intermediate and high QED peaks is retained in seed-based sampling. Generated molecules that fall between these two peaks might be particularly interesting candidates for evaluation, as such molecules still exhibit reasonably high drug-likeness but exist in a relatively sparse region of the training data and are therefore likely to offer more novelty. In any case, it seems that molecular seeds can efficiently guide the decoder toward generating new molecules with drug-like properties, even at high noise levels.

Overall, both seed-based and random sampling approaches were quite able to generate molecules with high to moderate QED without any deliberate external conditioning, which is an encouraging characteristic for generative small molecule design.

**Synthetic accessibility**   We now turn our focus to the synthetic accessibility (SA) of the generated molecules. The training set SA distribution has a maximum at ~2.1, with most active and inactive molecules falling below SA = 4. Most training molecules, therefore, are relatively easy to synthesize according to the SA measure. Random molecular generation (**rand0** and **rand1**) largely reproduced the training SA distributions, especially at low noise; the principal peak shifted toward an even lower mode of ~1.9 in the generative distributions, in part due to the lower average molecular size in generated molecules. SA distributions broadened at higher standard deviations (trending toward less synthesizable molecules), but maintained a prominent peak at an SA below the training mode. As the sampling standard deviation increases, molecules become smaller but more diverse; the reduction in size likely facilitates chemical synthesis, but the diversification might make synthesis challenging.

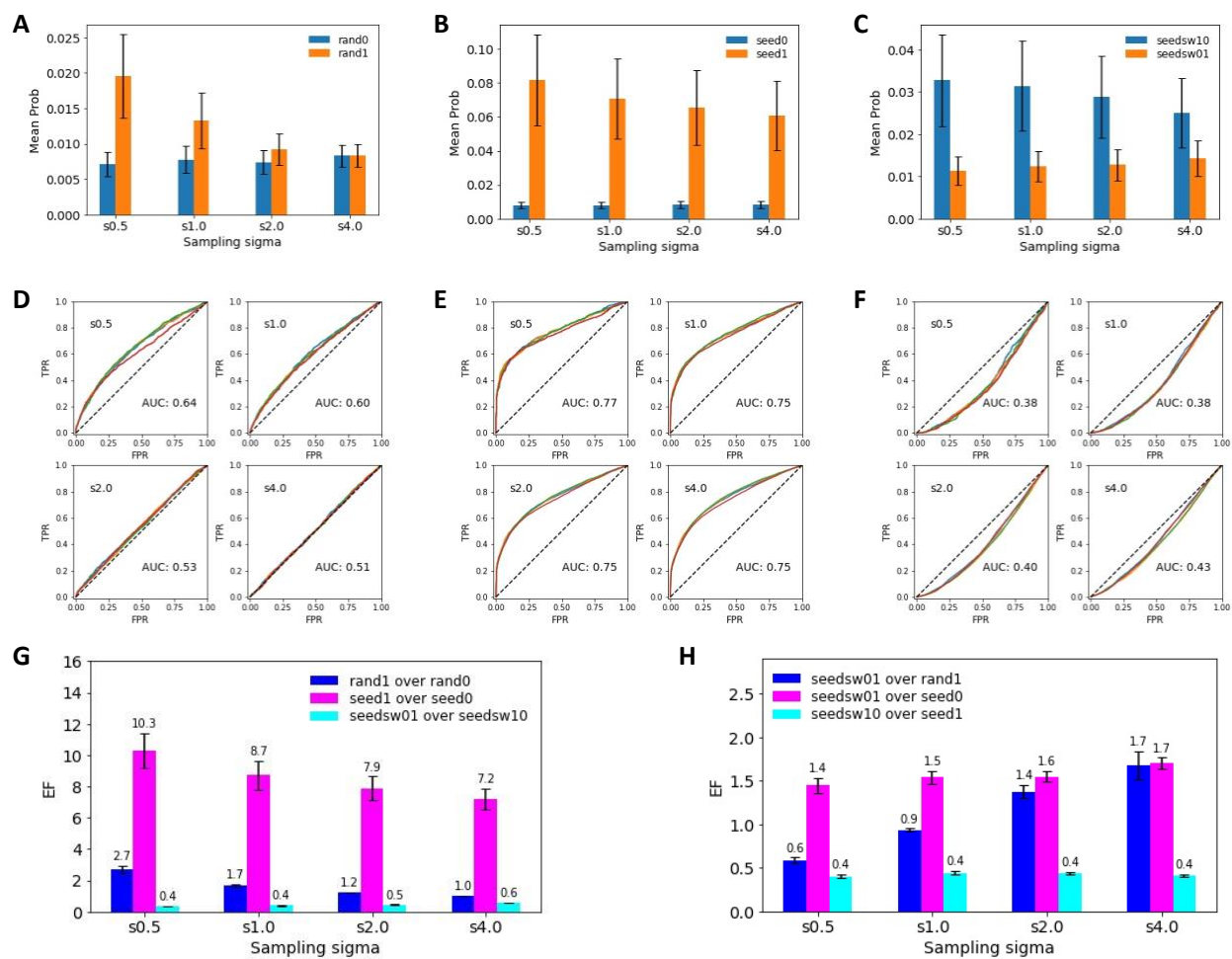

**Figure 3** | Conditional molecular generation for the dopamine D2 receptor. **A**, **B** & **C**, mean probabilities for generated molecules to be active from random, seed-based, and condition-swapped sampling, respectively. **D**, **E** & **F**, AUC and ROC for generated molecules from random, seed-based, and condition-swapped sampling, respectively. (Each colored line was obtained from four independently trained DL-QSAR models.) **G** & **H**, enrichment factors (EF) corresponding to each sampling condition and between various sampling conditions. EF is defined as the ratio of mean probabilities under comparison. EFs are computed between active and inactive conditions in each generation protocol (**G**) and condition-swapped and standard random or seed-based conditions (**H**).

Seed-based sampling also yielded SA distributions similar to those for the seeds themselves. Peaking at SA = ~2.0, the low-end sides of the generative distributions match the seed data particularly well; however, generated molecules tend to fall into the higher SA region more frequently (e.g., beyond SA = ~3.0, where seeds are seldom found). Nonetheless, the seed-based SA distributions seem less sensitive to the sampling noise level than those generated with random sampling, suggesting a trade-off between diversity and quality of generated molecules in terms of synthetic accessibility.

All told, our analysis of selected physical properties confirms that our generative model can effectively learn and reproduce such features without explicit conditioning, capturing properties that are buried in the high dimensional latent space, even with a different modality in expressing learned features between seed-based and random sampling. Differences in random and seed-based results suggest that either sampling method might be desirable, depending on the goals of the end user with respect to molecular divergence from training data.

**Conditional Generation Produces Predicted Biologically Active Molecules**

We have thus far examined the extent to which our generative model can reproduce unconditioned physical properties of training molecules. In general, patterns over physical properties were independent of activity condition. Despite this fact, can we generate molecules that are more likely to conform to a specified condition on biological activity? We next turn our focus on whether and to what extent an applied external condition can modulate generated molecules' biological activity. To assess activity of generated molecules, we first trained DL-QSAR models using data overlapping with the generative training set. These DL-QSAR models estimate the probability that a molecule is active against the given target.

**Random sampling** One of our most pressing questions was whether random sampling could produce molecules with activity with no seed molecules as a basis. In **Fig 3A**, we present the mean active probability for each activity group as a function of noise. As mentioned in the Methods section, activity was modulated by directly appending activity controlling bits, i.e., '10' for inactive (**rand0**) or '01' for active (**rand1**) at the end of a latent vector randomly sampled from the prior (i.e., $z \sim N(\mathbf{0}, s)$, where s = 0.5, 1.0, 2.0, or 4.0), producing conditional latent embeddings that can be fed into the decoder. **Fig 3 A&G** show that randomly sampled, active-conditioned molecules (**rand1**) have a higher probability of being active than inactive-conditioned molecules (**rand0**) at all standard deviations studied, although the ratio of activity probabilities between active and inactive conditions decreases as the noise level increases. **Fig 3A** shows that this trend can mostly be attributed to the probability changes in **rand1** rather than in **rand0**, which serves as a baseline for generating inactives with random sampling.

We also performed ROC analysis on the true positive (TPR) and false positive (FPR) rates between the active and inactive groups by sweeping the classification threshold, where TPR and FPR are defined by TP/(TP+FN) and FP/(TN+FP), respectively. In our analysis, the ground truth of each molecule was determined according to the applied activity condition in the latent vectors. For example, if a molecule were generated with the active-(or inactive-)condition, the molecule would be assigned as TP (or TN), regardless of any potential experimental label. In binary classification applications, a higher AUC typically implies better classification performance. In our case, however, higher AUC corresponds to better conditioning results with respect to the applied activity parameter.

Under this definition, we obtained an AUC of 0.64 at s0.5, suggesting that our conditioning procedure is successful to at least some extent at low noise levels. Though this AUC gradually decays to 0.51 at high noise (**Fig 3D**) in a manner consistent with the decay in mean activation probability profile (**Fig 3A**), our models are still quite capable of modulating activity without molecular seeds when noise is low.

These results do suggest, of course, that the efficacy of activity conditioning is highly sensitive to the sampling distance. As latent embeddings were sampled farther and farther from the prior mean (e.g., s4.0), the generated molecules became less and less recognizable as active molecules to the DL-QSAR model. While some molecules generated far from the training set might simply lie outside the scope of what a QSAR model can recognize, it is also reasonable to presume that fewer molecules, on average, will be active against an intended target as those generated molecules grow more diverse.

**Seed-based design** The DL-QSAR assessment for the seed-based sampling case (**Fig 3B**) shows a dramatic difference in mean activity probability when either active- and inactive-conditions are applied. The activation probabilities of active-conditioned molecules with active seeds (**seed1**) were higher than those of inactive-conditioned (**seed0**) molecules with inactive seeds by as much as 10.3 times (see the enrichment factor (EF) at s0.5), retaining an EF of 7.2 times even at the highest noise level (EF at s4.0) (see Fig 3G). The enrichment factor is defined as the ratio between mean active probabilities in comparison. Here, EF is calculated for the mean activation probability ratio between active and inactive conditioned molecules (i.e., $P_{rand1}/P_{rand0}$, $P_{seed1}/P_{seed0}$, and $P_{seed01}/P_{seed10}$). As with random sampling, the EF reduction as a function of increasing sampling standard deviation can mostly be attributed to a change in the probability of **seed1** rather in **seed0**. In subsequent ROC analysis, the highest AUC was 0.77 at s0.5, but still stays at 0.75 in other sampling conditions, implying a similar activity conditioning performance even at remote sampling distances.

Despite a qualitative similarity to random sampling, seed-based generation showed a clear quantitative difference in two respects. First, a much larger difference exists in the activation probabilities between seed- and inactive-conditioned molecules; second, that difference is still significant even when molecules are sampled from at high noise (e.g., s4.0). The seed-based approach is thus more efficient in generating molecules with specified activity, as long as one is able and willing to supply a specific activity-labeled seed to be used as the basis for generation.

These results provide two opposing insights about a seed's role in molecular generation. Seed molecules can serve as practical guides for searching chemically stable and functionally active molecules in a complex, high-dimensional latent space, even at a relatively long distance from the seed. On the other hand, a seed molecule likely has a strong and inherent memory effect that resists the control of latent vectors with activity conditions. In cases in which one wants generated molecules to diverge from known molecules yet retain activity (or perhaps inactivity, in the case of an off-target), this memory effect could be undesirable.

**Seed-Based Condition-Swapping Reveals Strong Memory Effect of Seeds on Molecular Activity in Generated Molecules**

To evaluate the extent of seed memory on resulting activity, we simulated "activation" and "deactivation" of molecules by swapping activity conditions from the known activities of designated seeds. Specifically, we conditioned latent vectors encoded from active seeds with inactive labels prior to decoding to produce the **seed10** molecule set (and vice versa for **seed01**. By comparing activity probabilities from condition-swapped with the regular seed-based results, we aimed to isolate the impact of seed memory on activity conditioning.

The mean activation probabilities for the condition-swapped generation cases (**Fig 3C**) implies that seed memory plays an important role in activity modulation. For example, at s0.5, the average activity probability for molecules generated using the inactive condition and active seeds (**seed10**) was about three times higher than that for molecules produced by conditioning inactive seeds with the active labels (**seed01**). Thus, even though we applied desired activities as conditions on the latent spaces, the inherent molecular features of seed molecules that underlie their activity were still reflected in the generated molecules.

Nonetheless, we found that the mean activation probability of molecules generated using active seeds and conditioned with inactive labels (**seed10**) was greatly reduced compared to that observed with active seeds and active labels (**seed1**) (**Fig 3 B&C**). Conversely, the activity probability of molecules generated using inactive seeds grew with the addition of active labels (**seed01**) compared to inactive labels (**seed0**). Moreover, the EF between the activity probabilities for the two condition sets (i.e., $P_{seed01}/P_{seed10}$) increased from 0.4 at s0.5 to 0.6 at s4.0 (**Fig 3G**). An EF less than one means that the deactivation of active seeds (**seed10**) still produces more active molecules than does activation of inactive seeds (**seed01**). However, the noise-dependence in the EF is noteworthy, because the activity probability during deactivation (**seed10**) decreases at higher standard deviation, whereas that of activation (**seed01**) increases (**Fig 3C**). This trend is in stark contrast to random or regular seed-based generation, where the activity probability of active-conditioned molecules decreased with increasing noise, while that of inactive labels remained roughly constant.

The extent of seed memory may explain the abnormal trend found in the condition-swapped generation. Indeed, sampling at high noise (e.g., s4.0) can annihilate the seed's inherent features that influence its biological activity. With such inherent features blurred, the external application of an activity condition potentially becomes impactful.

Seed memory can also be observed in the ROC analysis for generated molecules. For example, the AUC at s0.5 is estimated to be 0.38, a value significantly less than 0.5. From a classifier perspective, this means the classifier tends to predict more "negative" results (i.e., inactive) for true positive molecules (i.e., molecules generated with an "active" condition), but more "positive" results (i.e., active) for the true negative molecules (i.e., molecules generated with an "inactive" condition), demonstrating that the seed still exerts a noticeable effect in the opposite direction of the specified conditions. As the sampling noise increases, the AUC actually increases (reaching 0.43 at s4.0), suggesting the impact of the seed is diminished. As the AUC approaches 0.5, the model loses classification power and differences between classes become imperceptible. In the present context, the increase in AUC indicates compensation between seed memory and external conditioning effects.

To what extent, then, does direct conditioning of the latent space impact switching the resulting activity? To explore this question, we compared generated molecules' activity probabilities across different generation protocols (**Fig. 2H**). The enrichment factor for **seed01** over **rand1** increases from 0.6 at s0.5 to 1.7 at s4.0 (see blue bars in **Fig 3H**). This change shows that while the inactive seed holds strong influence at low noise (suppressing the active molecule generation probability to half that seen with random sampling), the seed's grip is loosened at high noise levels, allowing the external activity condition to have a greater relative influence on activating an inactive seed.

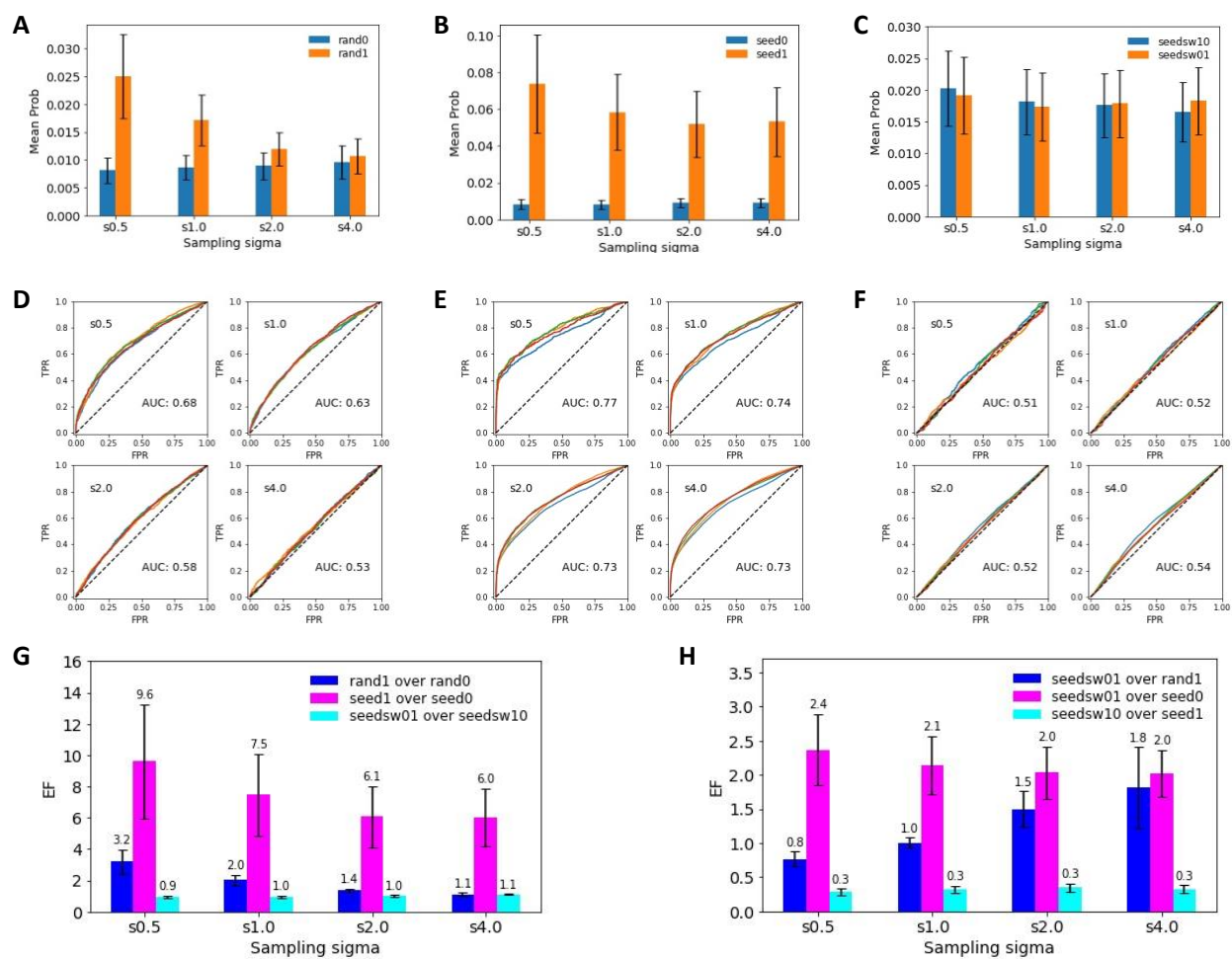

**Figure 4** | Validation for conditional molecular generation for D2 receptor with a DL-QSAR model trained on held-out data. **A**, **B** & **C**, mean probabilities for generated molecules to be active from random, seed-based, and condition-swapped sampling, respectively. **D**, **E** & **F**, AUC and ROC for generated molecules from random, seed-based, and condition-swapped sampling, respectively. (Each colored line was obtained from four independently trained DL-QSAR models.) **G** & **H**, enrichment factors (EF) corresponding to each sampling condition and between various sampling conditions. EF is defined as the ratio of mean probabilities under comparison. EFs are computed between active and inactive conditions in each generation protocol (**G**) and condition-swapped and standard random or seed-based conditions (**H**).

To more directly probe the effect of conditioning on predicted activity, we compared the mean activity probabilities between the two condition-swapped cases with those for molecules generated from only the initial seed (**seed01** vs. **seed0** and **seed10** vs. **seed1**). For the former, inactive seed case (see the orange bars in **Fig 3H**), the EF varied from 1.4 at s0.5 to 1.7 at s4.0, which means that the molecules were up to 70 % more likely to be active upon application of an active condition to an inactive seed. In the latter, active seed case (see green bars in **Fig 3H**), the EF stayed around 0.4 at all noise levels, implying that generated molecules were 60% less likely to be active upon application of an inactive condition to an active seed. The use of external conditions, therefore, certainly has an impact on activity predictions for generated molecules.

**Complementary Training Set Confirms Success of Activity-Conditioning Protocol**

Our results indicate that applying an external activity condition has a strong influence on determining the biological activity of generated molecules in all examined generation protocols. However, thus far, we have only evaluated generated molecules using DL-QSAR models trained on the same set of active molecules on which the generative model was built. Our initial analysis of generated molecules, therefore, was undoubtedly biased by the data used to train our evaluation model. The predictions made using that classifier are non-trivial, since only unique generated molecules (i.e., molecules not present in the training set) were assessed; nonetheless, that model still likely relies too overtly on similarities between generated molecules and the seeds on which they are based. To address this dataset bias issue, we built a complementary pair of generative and classification models trained on entirely separate subsets of the D2 dataset, allowing us to perform a more rigorous evaluation of our generative models (**Fig 4**). The DL-QSAR model yields an activity prediction AUC of 0.76 when applied to the generative training set, representing reasonable performance.

As shown in **Fig 4A&B**, active conditions applied in both random (**rand1**) or seed-based (**seed1**) protocols yielded more active molecules than did inactive conditions (**rand0** and **seed0**). In random sampling, the activity probability difference between the two activity conditions differed by a factor of up to 3.2, increased from the enrichment factor observed with the common training dataset (Fig. 3G/4G). In the seed-based case, the enrichment factor was slightly diminished relative to the common training set result, but still exhibited a dramatic difference of up to 9.7 times (Fig 4G). Decreases in these metrics are somewhat expected in the context of seed-based generation, considering that the complementary DL-QSAR model is now blind to seed molecules on which the generative model was trained. Nevertheless, the qualitive trends in activity in relation to the activity conditions still hold, suggesting that the generative model can learn essential features that result in active molecules, and that the independent classifier can relate those features back to activity.

Similarly, conditional AUCs seem to show improved results (e.g., 0.53~0.68 in the random sampling) or comparable (e.g., 0.73~0.77 in the seed-based) compared to those from models trained on the common dataset (**Fig 4D&E**). This observation suggests that the relative molecular distributions between the common and complementary datasets did not significantly change. However, since seed-related bias was removed from the complementary evaluation dataset, this DL-QSAR model allows for a more objective assessment of the impact of latent-space activity conditions.

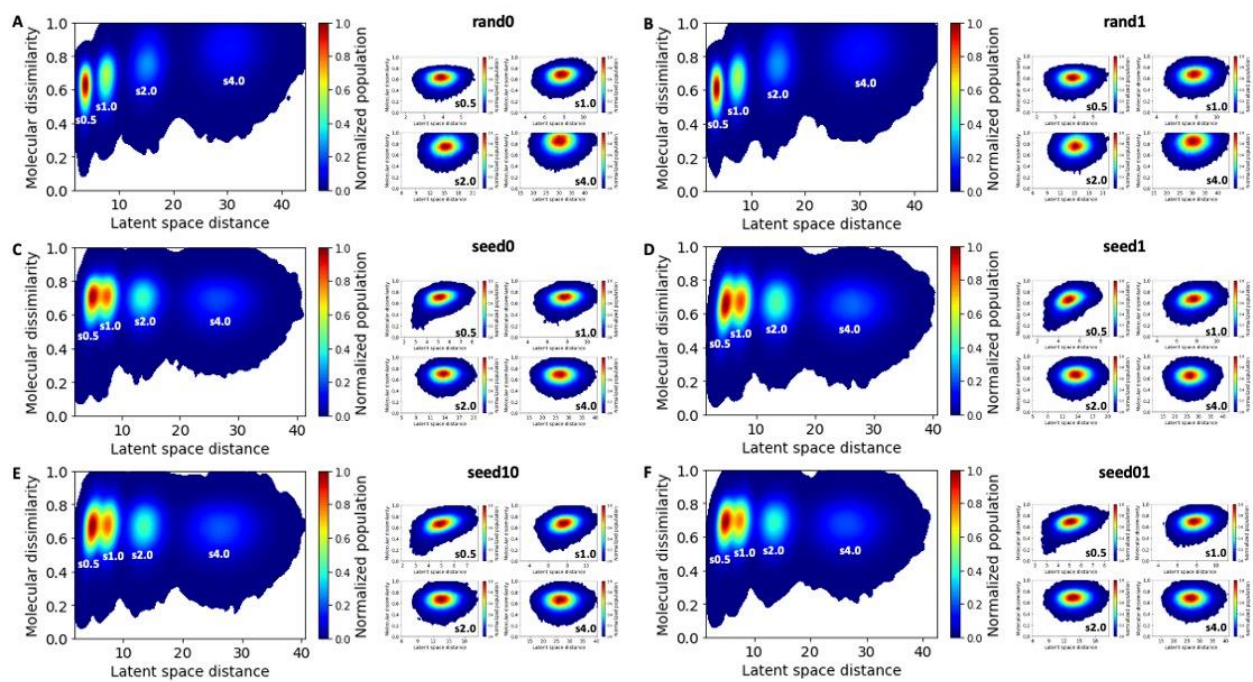

**Figure 5** | Relationship between molecular dissimilarity and latent space distance. **A** & **B**, random sampling with inactive and active conditions, respectively. **C** & **D**, seed-based sampling with inactive and active conditions, respectively. **E** & **F**, condition-swapped seed-based sampling for deactivation and activation, respectively.

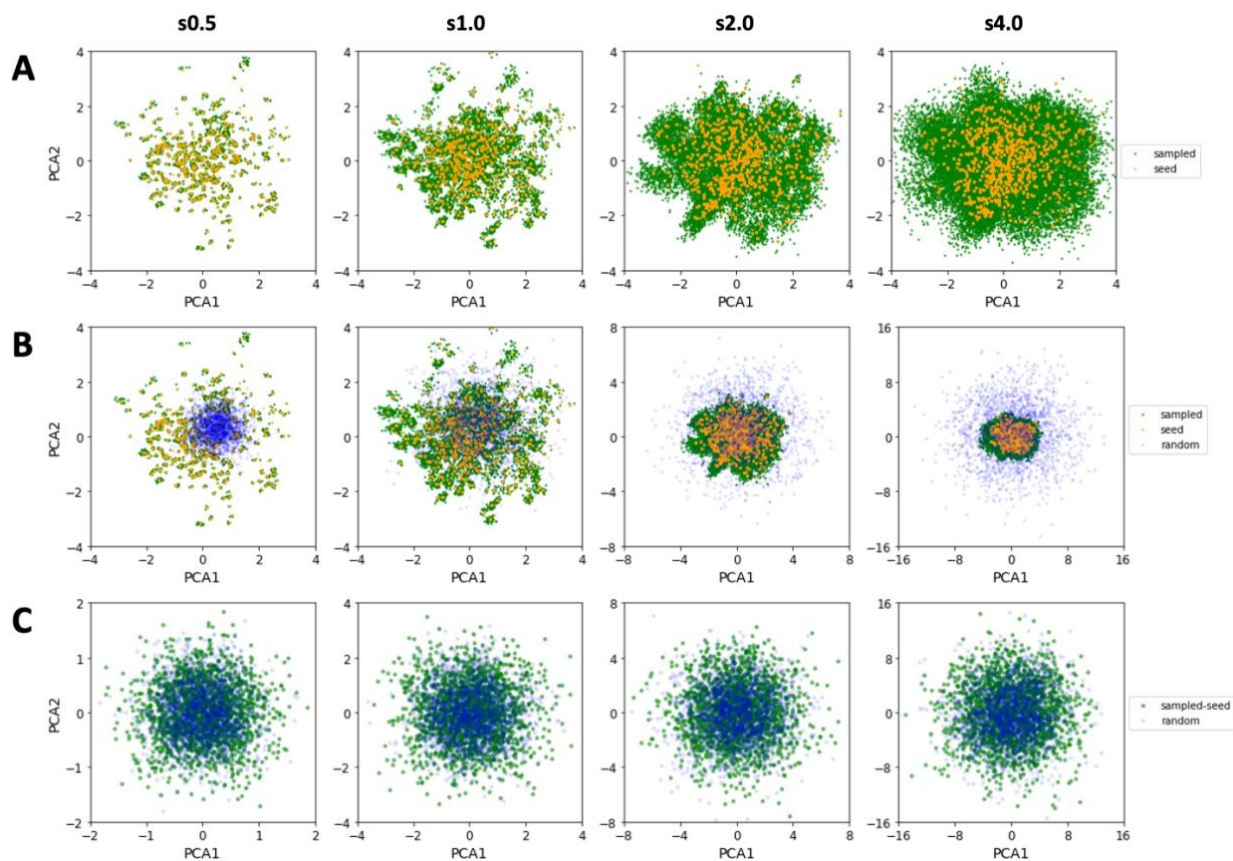

**Figure 6** | Principal component analysis of latent space vectors. **A**, PCA distributions of latent vectors (green) sampled from seeds (orange) as a function of sampling standard deviation. **B**, latent vectors (blue) sampled from a Gaussian prior at various sampling standard deviations superimposed on seed-based latent vectors. **C**, seed-based latent vectors (green) after removal of seed contributions superimposed on random generation results (blue).

To that end, we re-evaluated the extent of activation and deactivation that could be achieved through condition-swapped, seed-based generative protocols (i.e., **seed01** and **seed10**). We observed trends somewhat different from those seen previously. All noise levels yielded equivalent activity probabilities (EF ≈ 1) in both sets of condition swapping results (**Fig 4C&G**). Though our independent classifier can capture general structure-activity relationships, it does not make trivial connections between seed molecules and their labels; this explicit removal of seed-based bias seems to mimic the loss of seed dependence observed at high noise (cf. s4.0 of **Fig 3C**). In the absence of seed memory, the classifier is likely better able to make more objective predictions based on activity-relevant features, which perhaps explains why the mean activity probability predictions are constant with respect to the sampling noise level. Relatively speaking, however, molecules predicted to be active are still more enriched at high noise when inactive seeds are activated *via* conditioning: the activity probability of molecules generated from inactive seeds (**seed01**) grew from 0.8 to 1.8 times the mean from active-conditioned random sampling (**rand1**) with increasing noise (**Fig 4H**).

Compared to standard seed-based generation, the **seed01** protocol generated up to 2.4 times more active molecules than the inactive-condition on inactive seeds (**seed0**), indicating successful molecular activation *via* condition-swapping (**Fig 4H**). Likewise, the **seed10** protocol suppressed active molecule generation by about half (EF = ~0.3) relative to simple **seed1** generation (**Fig 4H**), suggesting that active seeds can also be deactivated through conditioning. Overall, these results from our rigorous, independent validation scheme reflect those from the common dataset: our conditional generative models perform reasonably well in producing molecules that conform to a specified biological activity.

**Molecular Similarity is Correlated with Latent-Space Similarity**

In previous sections, we showed that our conditional generative model can generate molecules with desired physical properties and biological activities. We now shift our analysis to the structure of the latent space itself. How do random and the seed-based sampling approaches differ with respect to exploring the latent space? What effects do different sampling protocols exert on the diversity of generated molecules? To address these questions, we investigated the relationship between latent space distance and molecular similarity. The latent space distance between two molecular embeddings was calculated using a simple Euclidian distance $\Delta z_\sigma^g(i,j) = d(z_\sigma^g(i), z_\sigma^g(j)) = \|z_\sigma^g(i) - z_\sigma^g(j)\|$ between two latent vectors $z_\sigma^g(i)$ and $z_\sigma^g(j)$ sampled at a standard deviation $\sigma$, where $g$ indicates the sampling protocol ($r$ = random; $s$ = seed-based). For random generation, ($z_\sigma^r(i)$ and $z_\sigma^r(j)$) randomly drawn from the prior mean as two independent samples $i$ and $j$; in seed-based generation, $z_\sigma^s(i)$ and $z_\sigma^s(j)$ are sampled around $z_0^s(i)$ and $z_0^s(j)$, latent vectors directly encoded from respective seeds $i$ and $j$. The molecular similarity, $s_\sigma^g(i,j)$, between molecules $m_\sigma^g(i)$ and $m_\sigma^g(j)$ generated from $z_\sigma^g(i)$ and $z_\sigma^g(j)$,, was computed using the Dice metric for Morgan fingerprints of radius 2. A value of $\sigma$ =0 refers to latent vectors for molecular seeds. At each sampling standard deviation, 100,000 unique molecule pairs were randomly sampled and subjected the distance and similarity calculation. Results are presented in the form normalized histograms, as shown in **Fig 5**. We illustrate similarity in the form of molecular dissimilarity, $\Delta s_\sigma^g(i,j) = 1 - s_\sigma^g(i,j)$, to facilitate a positive correlation with changes in latent space distance: in principle, the larger the latent space distance, the more dissimilar the selected molecular pair is.

Comparing random and the seed-based generative protocols reveals distinct trends in latent space with respect to sampling method. Random generation produced a proportional relationship between the latent space distance and the intermolecular dissimilarity, while seed-based sampling resulting in more or less constant dissimilarity as a function of latent space distance (**Table 1**).

**Table 1** | Mean intermolecular dissimilarities, $\overline{\Delta s_\sigma^g}$ between generated molecules.

| σ | Rand0 | Rand1 | Seed0 | Seed1 | Seed01 | Seed10 |
|---|---|---|---|---|---|---|
| 0.5 | 0.61 | 0.60 | 0.70 | 0.66 | 0.69 | 0.67 |
| 1.0 | 0.67 | 0.66 | 0.70 | 0.67 | 0.69 | 0.67 |
| 2.0 | 0.75 | 0.76 | 0.69 | 0.66 | 0.69 | 0.67 |
| 4.0 | 0.82 | 0.81 | 0.68 | 0.65 | 0.68 | 0.66 |

While no global trend was observed in the seed-based case, modest correlations between dissimilarity and latent space distance were observed at smaller standard deviations, e.g., 0.41(0.33) at s0.5 and 0.24(0.18) at s1.0 for **seed1**(**seed0**) (**Table 2**).

**Table 2** | Pearson correlation coefficients between intermolecular latent space distances and dissimilarity.

**a. Random molecular generation**

| σ | $\Delta z_\sigma^r$ vs. $\Delta s_\sigma^r$ | |
|---|---|---|
| | Rand1 | Rand0 |
| 0.5 | 0.09 | 0.09 |
| 1.0 | 0.13 | 0.16 |
| 2.0 | 0.11 | 0.13 |
| 4.0 | 0.10 | 0.09 |

**b. Seed-based molecular generation**

| σ | $\Delta z_\sigma^s$ vs. $\Delta s_\sigma^s$ | | $\Delta s_0^s$ vs. $\Delta s_\sigma^s$ | |
|---|---|---|---|---|
| | Seed1 | Seed0 | Seed1 | Seed0 |
| 0.5 | 0.41 | 0.33 | 0.70 | 0.59 |
| 1.0 | 0.23 | 0.18 | 0.66 | 0.57 |
| 2.0 | 0.08 | 0.06 | 0.61 | 0.54 |
| 4.0 | 0.04 | 0.03 | 0.53 | 0.50 |

**c. Condition-swapped seed-based generation**

| σ | $\Delta z_\sigma^s$ vs. $\Delta s_\sigma^s$ | | $\Delta s_0^s$ vs. $\Delta s_\sigma^s$ | |
|---|---|---|---|---|
| | Seed10 | Seed01 | Seed10 | Seed01 |
| 0.5 | 0.41 | 0.34 | 0.57 | 0.43 |
| 1.0 | 0.24 | 0.21 | 0.58 | 0.49 |
| 2.0 | 0.08 | 0.08 | 0.55 | 0.49 |
| 4.0 | 0.04 | 0.03 | 0.49 | 0.44 |

$\Delta s_0^s$: Intermolecular dissimilarities between input seeds.
$\Delta s_\sigma^g$: Intermolecular dissimilarities between generated molecules with a protocol $g$ (=random or seed) at s sampling $\sigma$.
$\Delta z_\sigma^g$: Intermolecular latent space distances between sampled latent vectors with a protocol $g$ at a sampling $\sigma$.

What brings about these discrepancies in molecular similarity when corresponding latent space distance distributions are functionally identical? To understand the mechanism driving these differences, we explored the structure of the latent space with a principal component analysis (PCA) of latent space vectors. To visualize the effect molecular seeds, we superimposed sampled random and seed-based latent vectors on a 2-dimensional PCA plane on top of the seed vector distribution. As shown in **Fig 6**, the random and seed-based distributions are quite different. Notably, latent vectors sampled during seed-based generation are tightly clustered around their reference seeds; while increasing noise does introduce more variation in sampled latent vectors, the underlying seed distribution remains evident even at the highest noise level (**Fig 6A**). Random generation, however, yields a simple, radially symmetric distribution of latent vectors that increases in extent as the noise level increases, spanning a much larger area than the seed-based distribution beyond s2.0 (**Fig 6B**).

It is important to note that the extent of randomness is the same in the seed-based and the random sampling protocols at a given sampling standard deviation, despite appearing much less extensive in the seed-based case. To confirm this point, we obtained a transformation matrix *via* PCA fitting for generated seed-based latent vectors after subtracting the contribution from each underlying seed (i.e., $\Delta z_i^s = z_i^s - z_i^{s,0}$). These seed-subtracted latent vectors ($\Delta z^s$'s) were then plotted alongside the random latent vectors ($z^r$'s) a 2-dimensional PCA plane. As shown in **Fig 6C**, the resulting distributions occupy the same space on the PCA plane and exhibit indistinguishable standard deviations. These plots illustrate the strong modulatory effect that seeds have in controlling noise introduced into the latent space.

**Molecular Dissimilarity is Proportionate to Latent Space Distance**

The PCA analysis provides several insights into differences between sampling protocols. Molecules generated *via* the seed-based protocol are tightly localized around their reference seeds, suggesting that a seed's innate features are translated into generated molecules even at high noise levels. Accordingly, seed-based sampling results in a latent subspace that mirrors the distribution of the seeds, explaining the constant intermolecular similarity as a function of latent space distance observed over all sampling conditions in **Fig 5 C&D**.

By contrast, random molecular generation is able to cover much broader swath of the latent space, and the extent of the sample subspace depends strongly on the sampling noise level. At s0.5 or s1.0, latent vectors are generally sampled within the area in which the seeds are located. This distributional overlap suggests that the features implicit to training molecules have been well encoded in the latent space through the variational inference on the Gaussian prior, as the favorable molecular properties (e.g., QED, SA, and activity) produced through random sampling further confirms. At high noise levels, the random search area extends well beyond latent subspace covered by the seeds, resulting in the generation of more diverse molecules that deviate from the mean characteristics of the training molecules.

Though molecules generated through the seed-based protocol do not exhibit a large similarity change when separated in latent space, latent space distance and similarity are still somewhat correlated when the sampling noise is low (**Fig 5C&D**). Interestingly, a similar analysis on seeds themselves (encoded without any Gaussian noise) reveals a linear relationship between the molecular similarity and the latent space distance (**Fig 7**).

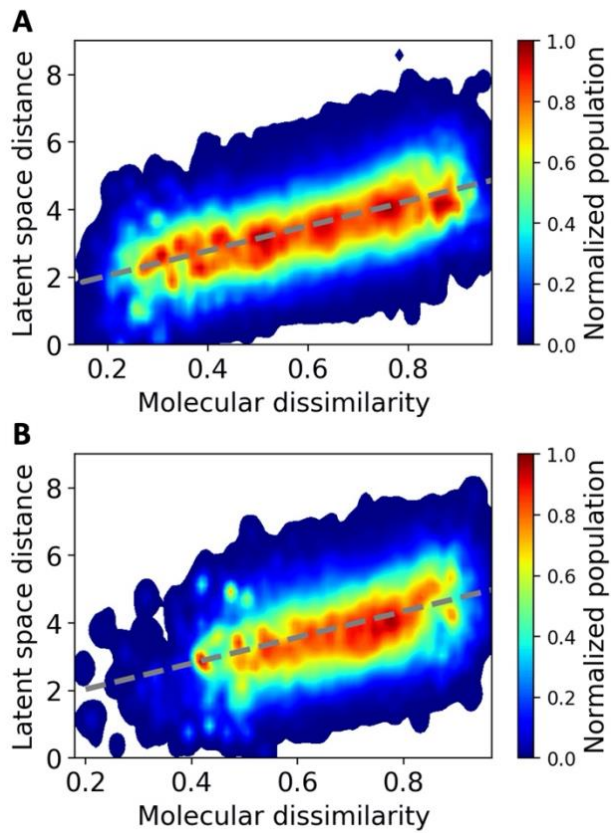

**Figure 7** | Intermolecular dissimilarity and latent space distance between the input seed molecules and their encoded latent space vectors (not sampled). **A** & **B**, active and inactive molecules, respectively.

We also examined the latent space distance-similarity relationship between generated molecules and their respective seeds (**Fig 8**). As the sampling standard deviation increases, the mean molecular dissimilarity between seeds and generated molecules also increases. In the **seed1**(**seed0**) protocol, for example, mean dissimilarity increases from 0.43(0.48) at s0.5 to 0.50(0.54) at s4.0 (inset of **Fig 8** and **Table 3)**. This general proportionality, between latent space distance and molecular dissimilarity is consistent with our PCA analysis, emphasizing that seeds serve as effective guides for generating seed-like molecules.

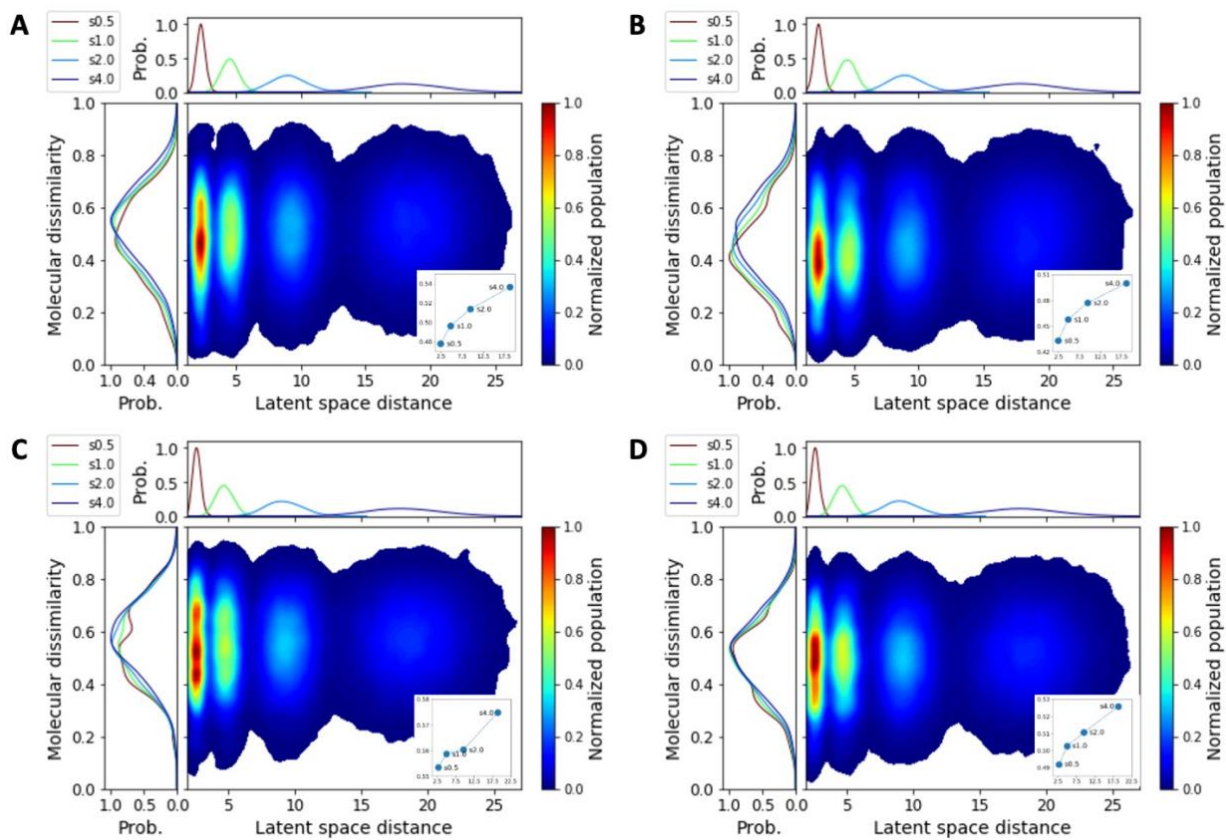

**Figure 8** | Intermolecular dissimilarity and intermolecular latent space distance between generated molecules and their reference seeds. **A** & **B**, distributions for seed0 and seed1 protocols, respectively. **C** & **D**, distributions for seed01 and seed10 protocols, respectively.

**Conditioning guides the generative network to search near the desired condition**

Lastly, we discuss the effect of condition swapping on the latent space distance and molecular similarity. Overall, the intermolecular similarity of generated molecules (**seed01** and **seed10**) to other generated molecules (**Fig 5 E&F**) or to their seeds (**Fig 8 C&D**) show patterns similar to those seen with standard seed-based generation (**seed0** and **seed1**) (see **Fig 5 C&D** and **Fig 8 A&B**). A few differences are noteworthy, as discussed in the following paragraphs.

First, compared to standard seed-based sampling, condition-swapped generation produced more dissimilar molecules from the designated seeds. Molecules generated *via* activation of inactive seeds (**seed01**) resulted in a mean dissimilarity at the lowest noise level that was higher than the dissimilarity seen at even the highest noise level for simple inactive seeding (**seed0**) (**Table 3**). A similar relationship was seen between the deactivation (**seed10**) and simple active seeding (**seed1**) protocols (**Table 3**). Comparing results at the highest noise level, the dissimilarity difference is significant for both condition-swapping protocols (**seed01** and **seed10**) with respect to the standard seeding schemes.

This change induced by the condition-swapping protocol gives an insight into how condition swapping works in the context activating or deactivating a seed. Activating or inactivating molecules with non-desired activity would require adding or subtracting features relevant to the desired activity that are presumably deficient in the reference seeds. Adding or subtracting such features would likely lead to generated molecules a longer distance away from the seeds, perhaps explaining the difference discussed above. Indeed, condition swapping diminished the correlation between intermolecular similarities of input seeds (i.e., $\Delta s_0^s(i,j)$) and corresponding output molecules (i.e., $\Delta s_\sigma^s(i,j)$). For example, the Pearson correlation coefficient 0.70(0.59) at s0.5 for **seed1**(**seed0**) was reduced to 0.57(0.43) for **seed10**(**seed01**) by swapping conditions prior to decoding (**Table 2**).

In contrast, we found that condition swapping drew the mean intermolecular dissimilarity closer to that seen with standard seed-based generation for the intended activity state (i.e., the end point of the condition swapping). While the mean dissimilarity produced from standard seed-based generation for inactive (active) molecules (Table 1), **seed0** (**seed1**) was 0.70 (0.66), the activation (inactivation) of those same seeds, **seed01** (**seed10**) shifted the mean value to 0.69 (0.67), which is closer to the observed dissimilarity value for standard seeding of the end state (0.66 (0.70) **seed1** (**seed0**)). Overall, the mean dissimilarity increases as a function of seeding protocol in the order of seed1 < seed10 < seed01 < seed0. These results highlight the tendency of generated molecules to be distributed according to the applied activity condition, regardless of the nature of underlying seeds.

**Table 3 |** Mean intermolecular dissimilarity for generated molecules from their reference seeds.

| σ | Seed0 | Seed1 | Seed01 | Seed10 |
|---|---|---|---|---|
| 0.5 | 0.48 | 0.43 | 0.55 | 0.49 |
| 1.0 | 0.50 | 0.46 | 0.56 | 0.50 |
| 2.0 | 0.51 | 0.48 | 0.56 | 0.51 |
| 4.0 | 0.54 | 0.50 | 0.57 | 0.53 |

## Conclusion

In this study, we extensively explored the impacts of training dataset and molecular seed bias on latent spaces of graph-based, activity-conditioned VAEs for small molecule generation. Employing a massive dataset labelled by activity against the dopamine D2 receptor, we trained generative models with explicit activity conditions and assessed those models with a separate DL-QSAR approach. We compared two molecule generation protocols, both at various levels of sampling noise: direct, random latent-space sampling and molecular seed-based sampling.

We first demonstrated that physical properties implicitly hidden in the training molecules can be effectively transferred in the generated molecules without explicit conditions in the seed-based protocol, albeit with less diversity in the resulting molecules relative to the random approach. We also showed that desired activity against a protein can be imposed on the generated molecules by modulating the latent space with binary activity conditions, even in the absence of seed molecules.

In condition swapping applications (in which seed molecules were activated or deactivated with an applied binary condition), we observed that the identities of seed molecules retained considerable influence on the application of activity conditions. To reduce the effect of seed memory on our models, we trained a separate generative model/DL-QSAR model pair on independent datasets, and we reevaluated the impact of seeds on the efficacy of external activity modulation. We found that activity conditioning remained successful in both random and seed-based generation protocols under this more rigorous evaluation metric.

PCA analysis revealed different sampling mechanics between the random and the seed-based protocols, especially with respect to the seed's strong influence in localizing samples in latent space. Comparisons of molecular similarity and latent space distance (on both seeds and generated molecules) highlighted a positive correlation between the input seeds and the encoder outputs, as desired and expected. Similarity analysis also provides insight into latent space sampling mechanisms: in particular, condition swapping guides the similarity distribution of the generated molecules toward the distributions sampled from seeds of the desired activity.

We hope the present study provides useful insights into the effects of training dataset and seed bias on molecule generation protocols, more rigorous strategies for evaluating generative models, and methods for optimizing latent space sampling/search algorithms for drug discovery applications of generative AI. While this model described in this work centers on 2D graph representations of molecules, many of the points discussed above likely apply to 1D SMILES representations of molecules or even 3D representations of protein-ligand complexes; comparisons of generation protocols across different molecular representations are certainly warranted in future work. Regardless, our current analysis indicates that the graph-based generative model presented above offers fine activity control and produces small molecules with favorable intrinsic properties, factors that should provide value to medicinal chemists.

† These authors contributed equally.

**Corresponding author**
* Email: sgkang@us.ibm.com